# НЕКОТОРЫЕ ОСОБЕННОСТИ ПЕРЕНОСА ИОННО-ИМПЛАНТИРОВАННОГО БОРА В УСЛОВИЯХ ПОДАВЛЕНИЯ СКОРОТЕЧНОЙ ДИФФУЗИИ

О.И. Величко, Е.А. Гундорина, В.В. Аксенов

Кафедра физики, Белорусский государственный университет информатики и радиоэлектроники, Беларусь, 220013 г.Минск, ул. П.Бровки, 6, тел. (+37529) 6998078, e-mail: velichkomail@gmail.com

Показано, что при термообработках слоев кремния, предварительно аморфизованных внедрением ионов германия и затем имплантированных ионами бора, перенос примеси вплоть до температуры 850 °C осуществляется посредством длиннопробежной миграции неравновесных межузельных атомов бора.

### Введение

Как известно, бор является основной примесью *p*-типа проводимости, используемой в технологии кремниевых интегральных микросхем. К сожалению, атомы бора обладают малой массой и большой подвижностью в кристаллах кремния. Вследствие малой массы ионов бора в имплантированных подложках даже при больших дозах внедрения не происходит образование аморфного слоя, а наоборот создается большое количество радиационных дефектов. Отсутствие аморфного слоя и наличие радиационных дефектов приводит к существенной скоротечной диффузии ионно-имплантированного бора. Все эти обстоятельства существенно усложняют задачу формирования мелких *p-n* переходов с высокими электрофизическими параметрами.

Для подавления скоротечной диффузии ионно-имплантированного бора в настоящее время широко используется метод введения бора в слой кремния, предварительно аморфизованный внедрением более тяжелых ионов германия, который, как и кремний, является элементом IV группы (смотри, например [1,2]). В результате твердофазной рекристаллизации аморфного слоя в области залегания имплантированного бора создается совершенная кристаллическая структура, практически не содержащая видимых посредством электронной микроскопии дефектов. Тем не менее, и в этом случае при последующих отжигах наблюдается скоротечная диффузия, хотя она носит совершенно иной характер и характеризуется меньшей интенсивностью. В работах [3,4] было показано, что при температурах обработки 800 °C и ниже скоротечная диффузия ионно-имплантированного бора осуществляется по механизму длиннопробежной миграции неравновесных межузельных атомов примеси. Согласно [4] генерация межузельных атомов бора может иметь место как в результате формирования или перестройки кластеров атомов примеси, так и в результате возникновения упругих напряжений в высоколегированной бором области или на границе слоя с высоким содержанием бора и слоя с высокой концентрацией германия в силу отличия атомных радиусов B и Ge от атомного радиуса Si. В то же время при температуре 900 °C и более высоких температурах [5,6] основную роль играет механизм образования, миграции и распада пар "атом бора — собственный межузельный атом", причем эти пары находятся в состоянии локального термодинамического равновесия с атомами бора в положении замещения и с неравновесными межузельными атомами кремния. Какой механизм диффузии доминирует в интервале температур 800 — 900 °C остается по-прежнему неясным.

### Цель работы

Целью данной работы является определение механизма диффузии ионно-имплантированного бора при температуре 850 °C, то есть в середине интервала, в котором была обнаружена смена микроскопического механизма диффузии.

### Особенности длиннопробежной миграции

В работах [7,8] были получены аналитические решения уравнения диффузии межузельных атомов примеси для случая непрерывной генерации этих межузельных атомов в пределах ионно-имплантированного слоя. Предполагалось, что скорость генерации этих неравновесных межузельных атомов примеси пропорциональна концентрации имплантированных атомов, то есть неоднородна в пределах данного имплантированного слоя. Как было показано в [7,8], протяженный "хвост" на профиле распределения концентрации примеси, сформированный в результате длиннопробежной миграции неравновесных межузельных атомов примеси, имеет вид прямой линии в случае логарифмического масштаба по оси концентраций. Указанная особенность формы профиля распределения примеси имеет место даже при небольшой протяженности "хвоста", сравнимой с характерным размером имплантированной области.

### Результаты расчетов

На Рис.1 представлен профиль распределения ионно-имплантированного бора после отжига в атмосфере азота в течение 60 секунд при температуре 850 °C, рассчитанный в предположении длиннопробежной миграции межузельных атомов примеси. Для сравнения здесь же приведены экспериментальные данные из [2]. В работе [2] имплантация ионов бора с дозой $1.0 \times 10^{15}$ см$^{-2}$ и энергией 1.5 кэВ осуществлялась в кремниевые подложки ориентации (100), имеющие глубокие карманы с высоким значением сопротивления, предварительно аморфизованные имплантацией ионов германия с энергией 12 кэВ и дозой $1.0 \times 10^{15}$ см$^{-2}$. Глубина аморфного слоя составляла 22 нм, то есть только незначительная часть имплантированного бора оказалась в кристаллический кремний.

При расчете были использованы следующие значения параметров имплантации и отжига: **параметры, задающие начальное распределения атомов бора:** $R_p$ =0.006 мкм (6.0 нм); $R_p$ = 0.00476 мкм (4.76 нм) и п**араметры, описывающие процесс межузельной диффузии:** средняя длина пробега межузельных атомов бора $l^{AI}$ = 0.0098 мкм (9.8 нм); усредненное по времени отжига значение скорости генерации неравновесных межузельных атомов примеси в максимуме распределения $g_m$ =1.56×10⁶ мкм⁻³с⁻¹. Как следует из проведенного расчета, приблизительно 3.74 % атомов имплантированного бора перешли в межузельное положение, что предоставило им возможность миграции в объем полупроводника, и затем перешли в положение замещения, что дало им возможность стать электрически активными.

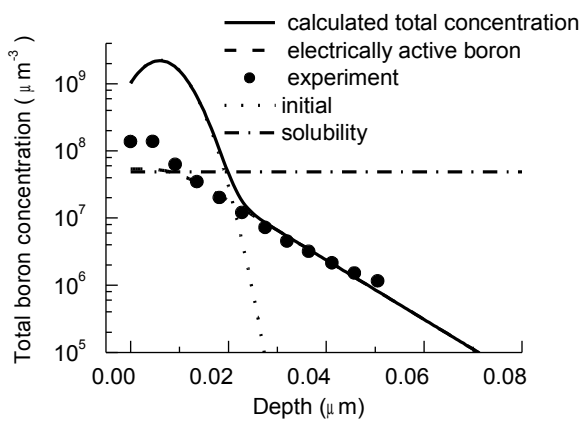

Рис.1. Рассчитанные распределения общей концентрации бора и концентрации атомов бора в положении замещения. Экспериментальные данные по распределению электрически активного бора после отжига при 850 °C в течение 60 секунд взяты из работы [2]. Штрих-пунктирная кривая — предел растворимости бора в кремнии

Как видно из Рис.1, результаты расчета хорошо согласуются с экспериментальными данными, причем протяженный "хвост" на профиле распределения концентрации бора действительно представляет прямую линию. Хорошее совпадение результатов проведенных расчетов с экспериментальными данными позволяет сделать вывод, что при температуре 850 °C перенос примеси в случае подавления скоротечной диффузии бора методом имплантации ионов в слой кремния, предварительно аморфизованный внедрением ионов германия, осуществляется посредством длиннопробежной миграции неравновесных межузельных атомов бора. Это означает, что смена микроскопического механизма переноса атомов бора с длиннопробежной миграции на диффузию посредством равновесных пар происходит в интервале температур 850 — 900 °C. Необходимо отметить, что в области высокой концентрации рассчитанный профиль распределения бора согласуется с пределом растворимости бора в кремнии. В то же время из представленных на Рис.1 экспериментальных данных следует, что в приповерхностной области размером ~ 8 нм концентрация бора в положении замещения превышает равновесный предел растворимости. На наш взгляд, это явление связано с влиянием поверхности на состояние дефектной подсистемы кристалла в приповерхностной области, что свою очередь влечет изменение предельной концентрации бора в положении замещения. Определенную роль на изменение предела растворимости могут также играть упругие напряжения, возникающие в высоколегированной приповерхностной области. Ответ на этот вопрос требует дальнейших исследований.

### Заключение

Проведено моделирование процесса перераспределения примеси при термообработках слоев кремния, предварительно аморфизованных внедрением ионов германия и затем имплантированных ионами бора. Показано, что перенос примеси вплоть до температуры 850 °C осуществляется посредством длиннопробежной миграции неравновесных межузельных атомов бора.

## SOME FEATURES OF THE TRANSPORT PROCESSES OF ION-IMPLANTED BORON UNDER CONDITIONS OF TRANSIENT ENHANCED DIFFUSION SUPPRESSION

Oleg Velichko, Alena Hundorina, Valerii Axenov

*Department of Physics, Belarusian State University of Informatics and Radioelectronics,*
*6, P.Brovki str., Minsk, 220013 Belarus*
*Tel. +375296998078, E-mail: velichkomail@gmail.com*

It has been shown that during thermal treatments of silicon layers preamorphized by germanium implantation and then implanted with boron ions the transport of impurity atoms occurs right up to a temperature of 850 °C due to migration of the nonequilibrium boron interstitials.